\documentclass[floatfix,amsmath,amssymb,superscriptaddress,longbibliography,twocolumn,showpacs]{revtex4-2}
\usepackage{amssymb}
\usepackage{amsmath}
\usepackage{bbold}
\usepackage{bm}

\usepackage{braket}
\usepackage[title]{appendix}
\usepackage{xcolor}
\usepackage{soul}
\usepackage{graphicx}
\usepackage{textcomp}
\usepackage{enumerate}
\usepackage{hyperref}
\usepackage{lipsum, babel}
\graphicspath{{../figura1/}{../figura2/}}
\usepackage{graphicx}

\newcommand*\diff{\mathop{}\!\mathrm{d}}

\usepackage{soul}

\usepackage{xcite}
\usepackage{xr}
\makeatletter

\begin{document}

\title{Predictive power of a Bayesian effective action for fully-connected one hidden layer neural networks in the proportional limit}
\author{P. Baglioni}
\affiliation{Dipartimento di Scienze Matematiche, Fisiche e Informatiche,
Universit\`a degli Studi di Parma, Parco Area delle Scienze, 7/A 43124 Parma, Italy}
\affiliation{INFN, Gruppo Collegato di Parma, Parco Area delle Scienze 7/A, 43124 Parma, Italy}
\author{R. Pacelli}
\affiliation{Dipartimento di Scienza Applicata e Tecnologia, Politecnico di Torino, 10129 Torino, Italy}
\affiliation{Artificial Intelligence Lab, Bocconi University, 20136 Milano, Italy}
\author{R. Aiudi}
\affiliation{Dipartimento di Scienze Matematiche, Fisiche e Informatiche,
Universit\`a degli Studi di Parma, Parco Area delle Scienze, 7/A 43124 Parma, Italy}
\affiliation{INFN, Gruppo Collegato di Parma, Parco Area delle Scienze 7/A, 43124 Parma, Italy}
\author{F. Di Renzo}
\affiliation{Dipartimento di Scienze Matematiche, Fisiche e Informatiche,
Universit\`a degli Studi di Parma, Parco Area delle Scienze, 7/A 43124 Parma, Italy}
\affiliation{INFN, Gruppo Collegato di Parma, Parco Area delle Scienze 7/A, 43124 Parma, Italy}
\author{A. Vezzani}
\affiliation{Istituto dei Materiali per l'Elettronica ed il Magnetismo (IMEM-CNR), Parco Area delle Scienze, 37/A-43124 Parma, Italy}
\affiliation{Dipartimento di Scienze Matematiche, Fisiche e Informatiche,
Universit\`a degli Studi di Parma, Parco Area delle Scienze, 7/A 43124 Parma, Italy}
\affiliation{INFN, Gruppo Collegato di Parma, Parco Area delle Scienze 7/A, 43124 Parma, Italy}
\author{R. Burioni}
\affiliation{Dipartimento di Scienze Matematiche, Fisiche e Informatiche,
Universit\`a degli Studi di Parma, Parco Area delle Scienze, 7/A 43124 Parma, Italy}
\affiliation{INFN, Gruppo Collegato di Parma, Parco Area delle Scienze 7/A, 43124 Parma, Italy}
\author{P. Rotondo}
\affiliation{Dipartimento di Scienze Matematiche, Fisiche e Informatiche,
Universit\`a degli Studi di Parma, Parco Area delle Scienze, 7/A 43124 Parma, Italy}
\date{\today}

\begin{abstract}
We perform accurate numerical experiments with fully-connected (FC) one-hidden layer neural networks trained with a discretized Langevin dynamics on the MNIST and CIFAR10 datasets. Our goal is to empirically determine the regimes of validity of a recently-derived Bayesian effective action for shallow architectures in the proportional limit. We explore the predictive power of the theory as a function of the parameters (the temperature $T$, the magnitude of the Gaussian priors $\lambda_1$, $\lambda_0$, the size of the hidden layer $N_1$ and the size of the training set $P$) by comparing the experimental and predicted generalization error. The very good agreement between the effective theory and the experiments represents an indication that global rescaling of the infinite-width kernel is a main physical mechanism for kernel renormalization in FC Bayesian standard-scaled shallow networks. 
\end{abstract}

\maketitle

\emph{Introduction---}  Understanding how deep learning \cite{GoodfellowBook} works, given the enormous success of its technological applications in the last decade \cite{modernCNN, transformers,resnet}, provides a timely challenge to theorists working in physics, mathematics and computer science \cite{zdeborova2020understanding, ZhangUnderstanding}. 

This need has prompted mathematical modeling of neural networks and their behavior in different learning scenarios, including (i) the so called infinite-width limit, that is informally defined as the regime where the size of each hidden layer $N_{\ell}$ ($\ell =1,\dots, L$, $L$ being the depth of the network) is much larger than the size of the training set $P$ \cite{Neal,NIPS1996_ae5e3ce4,g.2018gaussian,LeeGaussian,garriga-alonso2018deep,novak2019bayesian, JacotNTK,ChizatLazy,lee2019wide, pmlr-v119-bordelon20a, canatar2021}, (ii) standard perturbation theory in $1/N_\ell$ at finite size of the training set $P$ \cite{PDLT-2022, hanin2023random, zavatone-veth2021exact}; (iii) proportional limits where both the size of the layers $N_\ell$ and the depth $L$ are taken to infinity, keeping their ratio fixed \cite{hanin2022correlation}; (iv) analytically solvable toy models such as deep linear or globally-gated deep linear networks \cite{SompolinskyLinear, li2022globally, doi:10.1073/pnas.2301345120, PhysRevE.105.064118}; (v) different rescalings of weights and learning rate (wrt the standard Neural Tangent Kernel parametrization) to enhance feature learning effects in infinite-width networks \cite{doi:10.1073/pnas.1806579115, yang2020feature, NEURIPS2022_d027a5c9}. 

The aforementioned lines of research, while remaining of fundamental importance for the advance of our theoretical comprehension, mostly deal with settings that substantially deviate from the practically relevant ones, where the presence of a non-linear activation function is of utmost importance, and the number of trainset examples is usually larger than the average layer width $N_\ell$, but much smaller than the total number of parameters in the network \cite{resnet,vgg}. 

To the end of exploring more realistic settings (compared to state-of-the-art machine learning), progress has been made through a thermodynamic description to capture finite-width effects based on separation of scales \cite{NEURIPS2021_b24d2101,seroussi2023natcomm}, and, more recently, deriving an effective action for Bayesian fully-connected networks in the proportional limit where the size of the hidden layers $N_\ell$ is taken to infinity together with the size of the training set $P$ keeping the ratio $\alpha_\ell = P/N_\ell$ fixed  \cite{pacelli2023statistical}. In particular, for one hidden layer (1HL) networks, the result is found leveraging on a Gaussian equivalence \cite{mei2019, goldt2020gaussian, Gerace_2021,loureiro2021learning} informally justified by Breuer-Major central limit theorems \cite{BM, NourdinQuantitative, bardet2013}. Later, we have generalized this result to shallow architectures with one convolutional hidden layer \cite{aiudi2023}. 

From the physical viewpoint, the main consequence of the approach is that the infinite-width kernel of the architecture under consideration undergoes a renormalization in the proportional limit, firstly identified for deep linear models in \cite{SompolinskyLinear}, which can be entirely predicted by minimizing the corresponding effective action. Remarkably, the renormalization that takes place in fully-connected (FC) shallow networks is completely different from the one occurring in convolutional shallow networks \cite{aiudi2023}, pointing to a different, and more interesting, mechanism for feature learning at finite width. In one hidden layer fully-connected networks, the renormalized kernel that appears in the predictor's statistics is given by a global rescaling of the infinite-width one. 

Differently from infinite-width networks, where the equivalence with kernel learning is stated as a theorem, here we lack such a rigorous treatment. This means that it is of utmost importance to compare the predictions of the effective theory in the proportional limit with the outcome of actual numerical experiments.

In this manuscript, we perform accurate numerical experiments with FC one-hidden layer networks trained with a discretized Langevin dynamics on the MNIST and CIFAR10 datasets. By comparison with the generalization error predicted by the theory in the proportional limit, we establish that the Bayesian effective action captures the performance of the network consistently by varying the temperature $T$, the magnitude of the Gaussian priors, the size of the hidden layer and of the training set. 

\begin{figure}[t]
\includegraphics[width=0.45\textwidth]{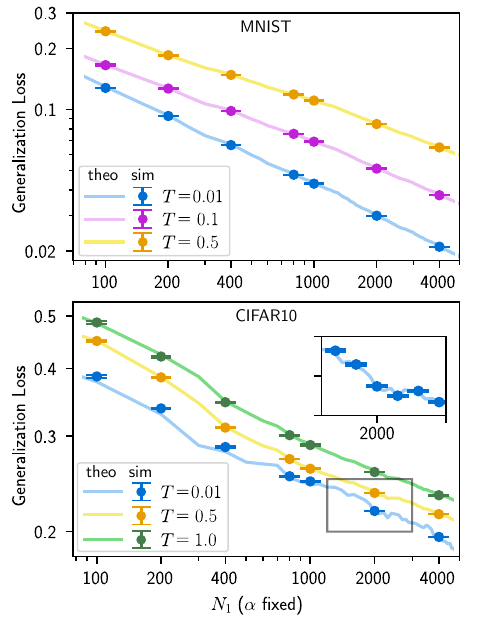}
\caption{The generalization loss of a fully-trained one-hidden layer network (with erf activation) is shown as a function of the size of the hidden layer $N_1$ at fixed $\alpha=P/N_1 = 1/4$ and $\lambda_0 = \lambda_1 = 1.0$ for two regression problems on the MNIST ($N_0 = 28 \times 28$) and (gray-scaled) CIFAR10 ($N_0 = 32 \times 32$) datasets (see supplemental material \cite{supp}). Solid lines represent the theoretical prediction computed from minimization of the effective action in Eq. \eqref{DNNeffectiveaction} at three different temperatures. Dots are the result of numerical experiments, where a discretized Langevin dynamics is employed to sample from the Bayesian posterior Gibbs ensemble (see \cite{paolo_github} for details). The inset shows a magnification of the learning curve at $T=0.01$ for the CIFAR10 dataset.}
\label{fig:1}
\end{figure}

Overall, our results provide a convincing indication that a main mechanism of kernel renormalization in  Bayesian FC one-hidden layer networks (with standard scaling) is given by global rescaling of the infinite-width kernel.

\emph{Predicted performance of a 1HL FC network in the proportional regime---} In the setting of \cite{pacelli2023statistical} - the proportional regime defined in the introduction and described in detail in the Supplemental material \cite{supp} - we have shown that the average test error of a Bayesian 1HL network on a new input $x^0 \in \mathbb R^{N_0}$ with scalar label $ y^0 $ (with mean-square error loss), can be expressed in the standard bias-variance decomposition as: 
\begin{align}
&\braket{\epsilon_\text{g}(x^0,y^0)} = (y^0-\Gamma)^2+\sigma^2, \label{eq:test_loss}\\
&\Gamma =\sum_{\mu\nu=1}^{P} \kappa^{\mathrm{(R)}}_\mu  \left(\frac{\mathbb{1}}{\beta}+K^{\mathrm{(R)}} ( \{\bar Q\}) \right)^{-1}_{\mu\nu}\; y_\nu, \label{eq:gamma}\\
& \sigma^2 = \kappa^{(\mathrm{R})}_{0} -\sum_{\mu\nu=1}^{P} \kappa^{(\mathrm{R})}_{\mu} \left(\frac{\mathbb{1}}{\beta} + K^{(\mathrm{R})}( \{\bar Q\})\right)^{-1}_{\mu\nu}\kappa^{(\mathrm{R})}_{\nu}.\label{eq:sigma2}
\end{align}
The renormalized kernel $K^{(\mathrm{R})}(\bar Q)$ is a $P \times P$ matrix that processes pairs of input training data $( x^\mu,  x^\nu)$ and it is given by: 
\begin{align}
K^{(\mathrm{R})}(\bar Q) = \frac{\bar Q}{\lambda_1} K (C)\,, \, \, \, \quad C_{\mu \nu} = \frac{1}{\lambda_0} \frac{ x^\mu \cdot  x^\nu}{N_0}\,,
\label{K_LQ}
\end{align} 
where $K$ is a non-linear operator (defined in the supplemental material \cite{supp}) that takes as input any (symmetric) 
matrix and computes a new matrix with the same shape.
Analogously to $K^{(\mathrm{R})}$, the renormalized quantities $\kappa_\mu^{(\mathrm{R})}$ and $\kappa_0^{(\mathrm{R})}$ process pairs of patterns comprising the new input $x^0$ (see \cite{supp} for more details). The hyperparameters $\lambda_1$ and $\lambda_0$ are scalar Gaussian priors for the weights of the last and first layer. 

The only free parameter on which the generalization error depends, that is $\bar Q$, is found minimizing the following effective action:
\begin{align}
S_{\textrm{FCN}}&=  -Q \bar Q + \log(1+Q) \nonumber \\
&+\frac{\alpha_1}{P}\text{Tr}\log \beta \left(  \frac{\mathbb{1}}{\beta} +K^{(\mathrm{R})}(\bar Q)\right) \notag\\
&+\frac{\alpha_1}{P} y^T \left( \frac{\mathbb{1}}{\beta} + K^{(\mathrm{R})}(\bar Q)\right)^{-1} y\,,
\label{DNNeffectiveaction}
\end{align}
where $y$ is the vector of train labels $y =(y^1,\dots,y^P)$. 
\begin{figure}[h!]
\includegraphics[width=0.45\textwidth]{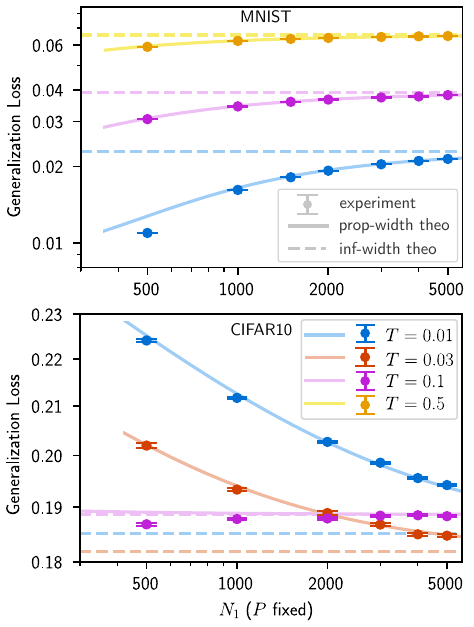}
\caption{The generalization loss of a fully-trained one-hidden layer network (with erf activation) is shown as a function of the size of the hidden layer $N_1$ at fixed size of the dataset $P = 1000$ for the same two regression problems in Fig. \ref{fig:1}. Solid lines represent the theoretical prediction computed from minimization of the effective action in Eq. \eqref{DNNeffectiveaction} at three different temperatures. Dots are the result of numerical experiments. Dashed lines correspond to the infinite-width prediction. Learning curves approach the infinite-width limit from below/above for the MNIST/CIFAR10 datasets in the entire range of temperatures probed (see also main text for a discussion on this point). Quantitative small deviations from the theory are observed at small $N_1$ (note that these are expected since the effective theory is derived in the proportional asymptotic limit).}
\label{fig:finite_t}
\end{figure}
It is worth noticing that the minimization of this effective action is straightforward if $\alpha_1 \to 0$, since one easily finds that $Q = 0$, $\bar Q = 1$ and recovers the well-known infinite-width limit Neural Network Gaussian Process (NNGP) kernel. 
As we show in the supplemental material \cite{supp}, in this proportional framework we can analytically compute another indicator of the network's performances, that is the test accuracy for a binary classification problem (i.e. one with binary labels $y^\mu = \pm 1$), namely: 
\begin{align}
\braket{\epsilon_{acc} (x^0, y^0)}= H\left(\frac{-\text{sign}(y^{0})\Gamma}{\sqrt{\sigma^2}}\right)
\label{eq:accuracy}
\end{align}
where $H(x) = 1/2\mathrm{Erfc}(x/\sqrt{2})$, and $\Gamma$ and $\sigma^2$ are the ones defined respectively in Eqs. \eqref{eq:gamma} and \eqref{eq:sigma2}. The observable in Eq. \eqref{eq:accuracy} can be interpreted as the expected percentage of trained 1HL Bayesian networks that are able to correctly classify the new example $(x^0, y^0)$. 

\begin{figure}[h!]
\includegraphics[width=0.45\textwidth]{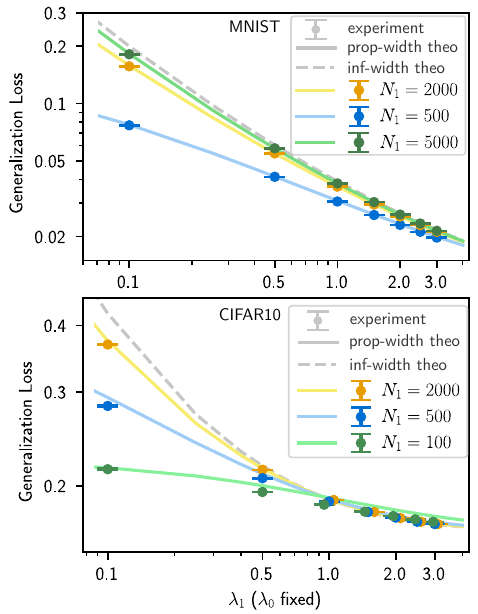}
\caption{The generalization loss is shown for 1HL networks with different size of the hidden layer as a function of the Gaussian prior over the last layer $\lambda_1$ (at temperature $T=0.1$ and $\lambda_0 = 1.0$ both for MNIST and CIFAR10). Solid lines represent the theoretical prediction computed from minimization of the effective action in Eq. \eqref{DNNeffectiveaction} at three different sizes of the hidden layer $N_1$. Dots are the result of numerical experiments. Dashed lines correspond to the infinite-width prediction. An improvement of the generalization performance is consistently observed across different network sizes, increasing the magnitude of the Gaussian prior $\lambda_1$. Whenever the three points at the same $\lambda_1$ overlap, we perform a small shift along the $\lambda_1$-axis on the two outer values to ease visualization, implying that the generalization loss was calculated at the same value of $\lambda_1$, related to the central point.}
\label{fig:lambda}
\end{figure}

\begin{figure}[h]
\includegraphics[width=0.45\textwidth]{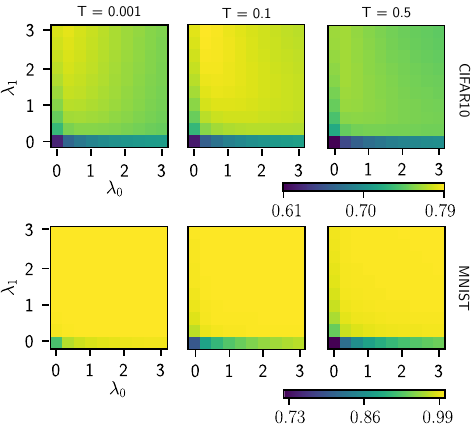}
\caption{The predicted test accuracy (Eq. \eqref{eq:accuracy}) is displayed as a function of the magnitude of the Gaussian priors over the first ($\lambda_0$) and last ($\lambda_1$) layer, for a 1HL network trained on two binary classification problems over the MNIST and CIFAR10 datasets ($P= N_1 = 1000$). Note that, while for the MNIST dataset increasing either $(\lambda_0$ or $\lambda_1)$ leads to better generalization, this is not the case for CIFAR10. Here it is always convenient to work at larger values of $\lambda_1$, but there exists an optimal intermediate value of the first layer Gaussian prior $\lambda_0$, which highlights an asymmetry in the role of regularization at different layers.}
\label{fig:boxplot}
\end{figure}

\emph{Numerical solution of the saddle-point equations at finite $T$---} The exact solution of the saddle-point equations deriving from the minimization of Eq. \eqref{DNNeffectiveaction} can be found only in the limit of zero temperature $T \to 0$ \cite{pacelli2023statistical}. At finite $T$, a numerical routine to find the global minima of Eq. \eqref{DNNeffectiveaction} must be implemented. Given the analytical expression of the renormalized kernel $K^{(\mathrm{R})}(\bar Q)$, an efficient way of doing this is to work in the basis where the infinite-width NNGP kernel $K^{(\mathrm{R})}(1)$ is diagonal (see \cite{supp} for more details and the online repository \cite{paolo_github} for a practical implementation of this numerical routine).

\emph{Langevin training---} To guarantee sampling from the canonical ensemble at temperature $T$ (which is fundamental to compare theory and numerical experiments), we choose to implement a Langevin dynamics on the weights of the network, collectively denoted as $\theta$, as was also done in the recent physics literature on Bayesian learning (see for instance Refs. \cite{SompolinskyLinear, seroussi2023natcomm, pacelli2023statistical}):
\begin{equation}
    \dot{\theta} = - \nabla_{\theta} \mathcal L (\theta)+ \sqrt{2T} \eta (t)\,,
    \label{Langevin}
\end{equation}
where $\mathcal L(\theta)$ is the regularized loss function defined in \cite{supp}. Note that, from the training perspective, this is just a gradient descent algorithm with an additional white noise $\eta(t) \sim \mathcal{N}(0,1)$ that introduces stochasticity in the learning dynamics. In the practical implementation of the algorithm, one introduces a discrete time step $\epsilon$ (or learning rate) in order to (numerically) integrate Eq. \eqref{Langevin}.

At least three critical issues must be carefully taken into account to obtain a genuine sampling from the Gibbs ensemble/Bayesian posterior: (i) one has to ensure that the dynamics has reached the stationary state, to avoid transient effects due to lack of thermalization; (ii) the integrated auto-correlation time should be estimated in the computation of uncertainties for physical observables; (iii) artifacts due to finite learning rate effects should be removed: one possible way is to extrapolate the results of simulations to $\epsilon \to 0$. These are well-studied issues in the community of gauge theory simulations \cite{Rothe2012, Gattringer2010}, from which we draw for the statistical analysis of the data (we discuss the numerical implementation in great detail in \cite{supp}).

\emph{Effect of temperature on generalization---} First, we explore the effect of temperature on generalization, analyzing the performance of 1HL networks on two regression problems with the MNIST and CIFAR10 datasets. In Fig. \ref{fig:1}, we display the predicted and experimental generalization performance as a function of the hidden layer size $N_1$, keeping fixed the ratio $\alpha = P/N_1$ to $\alpha=1/4$. Note that, such a prediction has no equivalent in the infinite width theory setting. In Fig. \ref{fig:finite_t}, we perform a similar analysis, but keeping fixed the number of examples $P$ in the training set. 

The effective theory in the proportional limit, based on the global rescaling of the infinite-width NNGP kernel, captures very accurately the outcome of numerical experiments at finite width, in the whole range of temperatures, sizes of the hidden layer and of the training set probed, except for quantitative deviations at small $N_1$, which are expected since the effective theory holds in the thermodynamic proportional limit. Lowering the temperature is always beneficial for generalization on the MNIST task, while the effect of thermal fluctuations on the CIFAR10 task is less trivial. In particular, in Fig. \ref{fig:finite_t} we can appreciate a crossing of the learning curves at the two different temperatures $T = 0.03$ and $T = 0.1$, which is again entirely predicted by the effective theory. We cannot exclude that the same phenomenology would take place for the MNIST task at a lower temperature that we are not probing with our numerical experiments. 

The fact that the optimal generalization performance for the CIFAR10 task is found at finite temperature is an indication that it is not always convenient to perfectly fit the training data. Since the temperature in the Bayesian setting is empirically related to the early-stopping time under gradient flow \cite{pmlr-v89-ali19a, NEURIPS2020_ad086f59}, it might be interesting to understand whether one can identify a dataset-dependent analytical criterion to determine the optimal temperature for improving the generalization performance of finite-width networks. 

\emph{Effect of Gaussian priors on generalization---} The temperature $T$ is not the only hyperparameter that we can fine-tune to improve the generalization performance. In principle, in our Bayesian 1HL network, we can also optimize the magnitude of the Gaussian priors of each layer, $\lambda_1$ and $\lambda_0$, which in the Langevin learning algorithm enter as $L^2$ regularization terms rescaled by the temperature. 

The results of this investigation are shown in Fig. \ref{fig:lambda} and Fig. \ref{fig:boxplot}. In the first plot, we examine the generalization loss at finite temperature $T=0.1$, as a function of the last layer Gaussian prior $\lambda_1$ (keeping fixed the first layer prior $\lambda_0$ to one) for three different sizes of the hidden layer. It turns out that it is always beneficial to run the training dynamics at large values of the last layer regularization $\lambda_1$. Note that this finding extends to finite-$T$ a more limited observation made in \cite{pacelli2023statistical} in the zero temperature limit. We also remark that the dependence of the generalization loss on the hidden layer size $N_1$ progressively disappears as long as the last layer regularization grows.  

The box plots in Fig. \ref{fig:boxplot} show the combined effect of fine-tuning both the Gaussian priors $\lambda_0$ and $\lambda_1$, on the predicted test accuracy for two binary classification tasks with a 1HL network with $N_1 = 1000$ hidden units at $\alpha=1$. While in the MNIST case, increasing the strength of regularization at each layer systematically leads to better test accuracies, for the CIFAR10 dataset we observe an optimal value of the first layer prior in the range $0 < \lambda_0 < 1$ (this is particularly evident in the central box plot at $T = 0.1$). On the contrary, it is always convenient to increase $\lambda_1$, as for the generalization loss in Fig. \ref{fig:lambda}.

This last point highlights an interesting asymmetry in the role of regularizations at each layer. We are planning to perform generalization experiments with state-of-the-art networks to understand whether the aforementioned observation may lead to potential improved performance in more realistic settings. 

\emph{Discussion: physical implications for feature learning---} The overall good agreement between the generalization performance predicted by the effective theory and our accurate numerical experiments with finite-width networks, represents a strong indirect evidence that global renormalization of the infinite-width NNGP kernel is a main mechanism for feature learning in Bayesian (standard-scaled) 1HL FC networks. Following the reasoning of our recent preprint \cite{aiudi2023}, this poses a limitation on the generalization capabilities of finite-width 1HL FC networks, a fact that finds empirical confirmation in a recent large-scale empirical study by the Google Brain group \cite{NEURIPS2020_ad086f59}.    

In the future, it would be interesting to extend the present comparison to (i) other observables of interest, such as the similarity matrix of internal representations (preliminarily studied in \cite{aiudi2023}), (ii) deeper FC architectures, (iii) shallow CNNs and (iv) shallow FC networks with multiple outputs, for which a Bayesian effective action is already available \cite{pacelli2023statistical}. 

\emph{Acknowledgements}.-- R.B. and P.R. are supported by $\#$NEXTGENERATIONEU (NGEU) and funded by the Ministry of University and Research (MUR), National Recovery and Resilience Plan (NRRP), project MNESYS (PE0000006) ``A Multiscale integrated approach to the study of the nervous system in health and disease” (DN. 1553 11.10.2022). 

\bibliography{biblio}

\clearpage
\onecolumngrid

\part*{Supplemental material}

\section{Setting of the learning problem}
We consider a supervised learning problem with training set $\mathcal D_P = \{x^\mu, y^\mu\}_{\mu=1}^P$, where each example $ x^\mu \in \mathbb R^{N_0}$ and the corresponding labels $y^\mu \in \mathbb R$. The architecture is a one-hidden-layer fully-connected neural network $f_{\theta} ( x)$ that implements the following function (with the standard scaling):
\begin{equation}
\label{eq:f_DNN}
f_{\theta} ( x) = \frac{1}{\sqrt{N_1}} \sum_{i_1=1}^{N_1} v_{i_1} \sigma \!\left[\frac{w_{i_1} \cdot x}{\sqrt{N_0}} \right]\,,
\end{equation}
where $\sigma$ is a non-linear pointwise activation function. We analyze regression problems with a quadratic regularized loss function:
\begin{align}
\mathcal L (\theta)  &= \frac{1}{2} \sum_{\mu=1}^P \left[y^\mu -f_{\theta}(x^{\mu})\right]^{2} + \mathcal L_{\textrm{reg}}(\theta)\,,\\
\mathcal L_{\textrm{reg}} (\theta) &=
   \frac{\lambda_1}{2\beta}\sum_{i_1} v_{i_1}^2+\frac{\lambda_0}{2\beta}\sum_{i_0 i_1} w_{i_0 i_1}^2.
\label{loss},
\end{align}
where $\beta =1/T$ is the inverse temperature parameter. 

As a standard practice in statistical mechanics of deep learning, we define the partition function of the problem as:
\begin{equation}
Z = \int \!\mathcal D \theta \,e^{-\beta \mathcal L (\theta)}\,.
\label{defpartition}
\end{equation}
where the symbol $\int \mathcal D \theta$ indicates the collective integration over the weights of the network, $\theta = \{ w , v  \}$. 
This choice enforces minimization of the training error for $\beta \to \infty$. Scaling $\mathcal L_{\textrm{reg}}(\theta)$ by $1/\beta$ has a natural Bayesian learning interpretation: the Gibbs probability $P_{\beta} (\theta)=Z^{-1} e^{-\beta \mathcal L (\theta)}$ associated with the partition function in equation \eqref{defpartition} is the posterior distribution of the weights after training, whereas the Gaussian regularization is a prior equivalent to assuming that weights at initialization have been drawn from a Gaussian distribution. In this framework, the expected value of a generic observable  $O(\theta)$ is computed over the posterior distribution of the weights as:
\begin{equation}
\braket{O(\theta)} = \int \!\mathcal D \theta \, O(\theta) \frac{e^{-\beta \mathcal L(\theta)}}{Z}\,.
\label{supp:eq:observable}
\end{equation}
In particular, the average test error over a new (unseen) example $(\mathbf{x}^0, y^0)$ is given by:
\begin{equation}
\braket{\epsilon_{\textrm g} (\mathbf{x}^0, y^0)} = \int \!\mathcal D \theta \, [y^0- f_{\textrm{DNN}}(\mathbf{x}^0)]^2 \frac{e^{-\beta \mathcal L(\theta)}}{Z}\,.
\label{eq:err_g_def}
\end{equation}.

\section{NNGP Kernel function and its renormalization in the proportional regime}
The non-linear kernel operator $K$, found extensively in the literature \cite{Neal, LeeGaussian, lee2019wide} and in the present work, takes as input any (symmetric) $P \times P$ matrix, that in this specific case is the dataset correlation matrix $C$, and computes a new matrix in the following way:
\begin{align}
\label{eq:K_munu}
K_{\mu\nu} (C) &= \int d^2 t\, \mathcal N_{t} \left(0,  \tilde C_{\mu\nu}\right) \sigma (t_1) \sigma(t_2)\,, \\
\tilde C_{\mu\nu} &= \begin{pmatrix}
C_{\mu\mu} & C_{\mu\nu} \\
C_{\mu\nu} & C_{\nu\nu}
\end{pmatrix} \,, \qquad C_{\mu \nu} = \frac{x^\mu \cdot x^\nu}{\lambda_0 N_0}
\end{align}
where we are denoting (here and in the following) normalized Gaussians as:
\begin{equation}
    \mathcal N_x \left( m, \Sigma \right) \equiv \frac{ \textrm{exp} \left( -\frac{1}{2}(x-m)^\top \Sigma^{-1} (x-m)\right) }{\sqrt{\mathrm{det}2\pi\Sigma }}.
\end{equation}
We recall the definition of the additional kernel integrals that depend on the new example $x^0$, that is $\kappa_\mu$, where the index $\mu = 0, \ldots ,P $ covers both the dataset and the new test element $x^0$:
\begin{align}
&\kappa_\mu = \int \diff^2 t \, \mathcal{N}_t \left(0, \tilde C_\mu \right) \sigma(t_1) \sigma (t_2)\,, \label{kmu} \\
&\tilde C_\mu = \begin{pmatrix} C_{\mu\mu} & C_{\mu0}\\ C_{\mu0} & C_{00}\end{pmatrix}\,, \, \, C_{\mu 0} =  \frac{x^\mu \cdot x^0}{\lambda_0 N_0} \,,
\end{align}
Note that $K$ can also be understood as a real-valued function of two variables $K(x^\mu, x^\nu) \equiv K_{\mu\nu}$. With this notation we have that $\kappa_\mu \equiv K(x^\mu, x^0)$. The aforementioned quantities completely define the output and generalization capabilities of networks in the infinite-width limit. In the proportional regime, on the other hand, the kernel matrix (and vectors) undergo a global (scalar) renormalization by the order parameter $\bar Q$ (see also \cite{pacelli2023statistical}), namely: 
\begin{align}
    & \left[ K^{\mathrm{R}} \right]_{\mu\nu} = \frac{ \bar Q}{ \lambda_1} K(x^\mu,x^\nu) =\frac{ \bar Q}{ \lambda_1} K_{\mu \nu}  \\
    & \kappa^{(\mathrm{R})}_\mu = \frac{ \bar Q}{ \lambda_1} K(x^\mu,x^0) =\frac{ \bar Q}{ \lambda_1} \kappa_{\mu} 
\end{align}
These renormalized quantities incorporate the information contained in the dataset $\mathcal D$, mediated by the parameter $\bar Q$ that optimizes the action in Eq \eqref{DNNeffectiveaction}, and completely define the network's behavior in the proportional regime, as shown by the expressions of the test loss and test accuracy.

\section{Average accuracy }
In this section we show how to compute the average accuracy on a
new test example $( x^{0},y^{0})$. The observable we
are interested to reads:
\begin{align}
\epsilon_{acc} & = \left\langle\Theta\left(y^{0}f_\theta(x^0)\right)\right\rangle,
\end{align}
where $\Theta(x)$ is the  Heaviside step function, and the average is over the Gibbs ensemble associated with the regularized mean-square error loss $\mathcal L (\theta)$ (see Eq \ref{supp:eq:observable}). Note that this observable is defined for classification problems, therefore we need the outputs to be binary $y^\mu = \pm 1 \, \, \forall \mu$. We have:
\begin{align}
\epsilon_{acc}&=\frac{1}{Z} \int d\theta \, \Theta\left(y^{0}f_\theta(x^0)\right)e^{-\beta\mathcal{L}\left(\lbrace\theta\rbrace\right)} \nonumber \\
 & =\frac{1}{Z}\int\prod_{\,i_{0}\,i_{1}}dw_{i_{1}i_{0}}\prod_{\,i_{1}}dv_{i_{1}}\ \theta\left(y^{0}f_\theta(x^0)\right)e^{-\frac{\beta}{2}\sum_{\mu}\left(y^{\mu}-f_{\theta}(x^{\mu})\right)^{2}-\frac{\lambda_1}{2}\sum_{i_1} v_{i_1}^2-\frac{\lambda_0}{2}\sum_{i_0 i_1} w_{i_0 i_1}^2} 
\end{align}
To make progress in the calculation, one needs to introduce two sets of Dirac deltas (one for the pre-activations and one for the network outputs), both for the trainset elements and for the new example: 
\begin{align}
\prod_{i_{1}\,\mu\,}\delta\left(h_{i_{1}}^{\mu\,}-\frac{1}{\sqrt{N_{0}}}\sum_{i_{0}=1}^{N_{0}}w_{i_{1}i_{0}}x_{i_{0}}^{\mu}\right)\qquad & \prod_{\mu\,}\delta\left(s^{\mu\,}-\frac{1}{\sqrt{N_{1}}}\sum_{i_{1}=1}^{N_{1}}v_{i_{1}}\sigma\left(h_{i_{1}}^{\mu\,}\right)\right)\\
\prod_{i_{1}}\delta\left(h_{i_{1}}^{0}-\frac{1}{\sqrt{N_{0}}}\sum_{i_{0}=1}^{N_{1}}w_{i_{1}i_{0}}x_{i_{0}}^{0}\right)\qquad & \delta\left(s^{0}-\frac{1}{\sqrt{N_{1}}}\sum_{i_{1}=1}^{N_{1}}v_{i_{1}}\sigma\left(h_{i_{1}}^{0}\right)\right).
\end{align}
When it is not specified, the index $\mu$ runs only on the trainset $\mu=1\ldots P$. Expressing the previous quantities in their exponential representation - through the introduction of an equal number of conjugate variables - allows Gaussian integration over the weights and factorization over the hidden-layer node index $i_1$.  The average accuracy then reads:
\begin{align}
\epsilon_{acc} & =\frac{1}{Z}\int\prod_{\mu=0}d^{2}s^{\mu}\Theta\left(y^{0}s^0\right)e^{-\frac{1}{2}\sum_{\mu}\left(y^{\mu}-s^{\mu}\right)^{2}+i\sum_{\mu=0}s^{\mu}\bar{s}^{\mu}} \nonumber \\
 & \qquad\times\left[\int\prod_{\mu=0}d^{2}h^{\mu}e^{i\sum_{\mu=0}h^{\mu}\bar{h}^{\mu}} e^{-\frac{1}{2}\sum_{\mu\nu=0}\bar{h}^{\mu}C^{\mu\nu}\bar{h}^{\nu}-\frac{1}{2\lambda_{1}N_{1}}\sum_{\mu\nu=0}\bar{s}^{\mu}\sigma(h^{\mu})\bar{s}^{\nu}\sigma(h^{\nu})}\right]^{N_{1}}.
\end{align}
Note that the covariance matrix $C$ here is extended to the new example $x^0$, becoming $P+1$-dimensional symmetric matrix with the extra row (and column) given by the vector $((x^\mu \cdot x^0)/\lambda_0 N_0, \,  x^\mu=0 \ldots P)$. Following the same steps of \cite{pacelli2023statistical}, we obtain the following expression for the generalization accuracy:
\begin{align}
&\epsilon_{acc}  =\frac{1}{Z}\int\prod_{\mu=0}d^{2}s^{\mu}\Theta\left(y^{0}s^0\right)  \exp \left[ -\frac{\beta}{2}\sum_{\mu}\left(y^{\mu}-s^{\mu}\right)^{2}+i\sum_{\mu=0}s^{\mu}\bar{s}^{\mu}-\frac{N_{1}}{2}\log \left( 1+Q(\bar s) \right)\right] \\
& \qquad Q(\bar{s}) =\frac{1}{\lambda_{1}N_{1}}\sum_{\mu\nu=0}\bar{s}^{\mu}K_{\mu\nu}\bar{s}^{\nu}
\end{align}
%
%
Introducing one more Delta function for the quantity $Q(\bar s)$:
\begin{equation}
 \delta\left(Q^{}-\frac{1}{\lambda_{1}N_{1}}\sum_{\mu\nu=0}\bar{s}^{\mu}K_{\mu\nu}^{}\bar{s}^{\nu}\right),
\end{equation}
and expressing it in its standard exponential form, we have:
\begin{align}
&\epsilon_{acc}=  \frac{1}{Z}\int dQ d \bar Q\,\Theta\left(y^{0}s^0\right)e^{\frac{N_{1}}{2}Q\bar{Q}-\frac{N_{1}}{2}\text{log}(1+Q)} \nonumber \\
 & \qquad \times\int\prod_{\mu=0}d^{2}s^{\mu}e^{i\sum_{\mu=0}s^{\mu}\bar{s}^{\mu}-\frac{1}{2}\sum_{\mu}\left(y^{\mu}-s^{\mu}\right)^{2}-\frac{1}{2 N_1 \lambda_{1}}\sum_{\mu\nu=0}\bar{s}^{\mu}\bar{Q}K_{\mu\nu}^{}\bar{s}^{\nu}}
\end{align}
Isolating the renormalized kernels $ K^{\mathrm{R}}$ and $ \kappa^{(\mathrm{R})}$, defined in the previous section, we can separate the contribution of the trainset elements from those of the test example $x^{0}$ and rewrite:
\begin{align}
\epsilon_{acc}&=  \frac{1}{Z}\int dQ d \bar Q e^{\frac{N_{1}}{2}Q\bar{Q}-\frac{N_{1}}{2}\text{log}(1+Q)} \int d^{2}s^{0}\Theta\left(y^{0}s^0\right)e^{is^{\mu0}\bar{s}^{\mu0}-\frac{1}{2} \kappa^{(\mathrm{R})}_0(\bar{s}^{0})^2} \nonumber \\
 & \qquad \times\int\prod_{\mu=1}d^{2}s^{\mu}e^{i\sum_{\mu}s^{\mu}\bar{s}^{\mu}-\frac{\beta}{2}\sum_{\mu}\left(y^{\mu}-s^{\mu}\right)^{2}-\frac{1}{2}\sum_{\mu\nu}\bar{s}^{\mu}\left[ K^{(\mathrm{R})} \right]_{\mu\nu}\bar{s}^{\nu}-\bar s_{0}\sum_{\mu}\kappa^{(\mathrm{R})}_{\mu}\bar{s}^{\mu}}
\end{align}
We now can perform respectively the integrations over $s^{\mu}$ with $\mu\neq0$, and all the $\bar{s}^{\mu}$, to obtain:
\begin{align}
\epsilon_{acc}= & \frac{1}{Z}\int dQ d \bar Q\, e^{\frac{N_{1}}{2}Q\bar{Q}-\frac{N_{1}}{2}\text{log}(1+Q)-\frac{1}{2}\text{log\,det}\left[ K^{(\mathrm{R})} +\mathbb{\frac{1}{\beta}}\right]-\sum_{\beta\mu\nu}y^{\nu}\left[  K^{(\mathrm{R})} +\mathbb{\frac{1}{\beta}}\right]^{-1}_{\mu\nu}y^{\mu}} \nonumber \\
& \qquad \times \int ds^{0} \frac{\Theta \left( y^{0}s^0 \right)}{\sqrt{2 \pi \sigma^2}} \exp \left[-\frac{(s_0 - \Gamma(\bar Q))^2}{2\sigma^2 (\bar Q)} \right]
\end{align}
where we have defined 
\begin{align}
\sigma^2 (\bar Q)  =\kappa^{(\mathrm{R})}_0-\sum_{\mu\nu} \kappa^{(\mathrm{R})}_\mu\left[K^{(\mathrm{R})}+\mathbb{\frac{1}{\beta}}\right]^{-1}_{\mu\nu} \kappa^{(\mathrm{R})}_\nu, \qquad \quad \Gamma (\bar Q) =\sum_{\mu\nu}y^{\mu} \left[ K^{(\mathrm{R})} +\mathbb{\frac{1}{\beta}}\right]^{-1}_{\mu\nu} \kappa^{(\mathrm{R})}_\nu,
\label{eq:bias_vas_replicated}
\end{align}
highlighting the dependence of mean and variance on the parameter $\bar Q$, that we can now drop to ease the notation. The accuracy reads:
\begin{equation}
\epsilon_{acc} =  \frac{1}{Z}\int dQ d \bar Q\, e^{-\frac{N_{1}}{2}S(Q,\bar{Q})}H\left(\frac{-\text{sign}(y^{0})\Gamma }{\sqrt{\sigma^2}}\right),
\end{equation}
where the function $H(x) = 1/2\mathrm{Erfc}(x/\sqrt{2})$. Note that the action $S(Q,\bar{Q})$ is the same obtained for the partition function $Z$,
therefore we have:
\begin{equation}
\epsilon_{acc}=H\left(\frac{-\text{sign}(y^{0})\Gamma}{\sqrt{\sigma^2}}\right)\mid_{Q^*, \bar Q^*}
\end{equation}
that has to be evaluated at the saddle point solution $Q^{*},\bar{Q}^{*}$ of the effective action
defined in \cite{pacelli2023statistical} and in the main text of this work.

\section{Monte Carlo training of Neural Networks}

\begin{figure}[t]
\includegraphics[width=.98\textwidth]{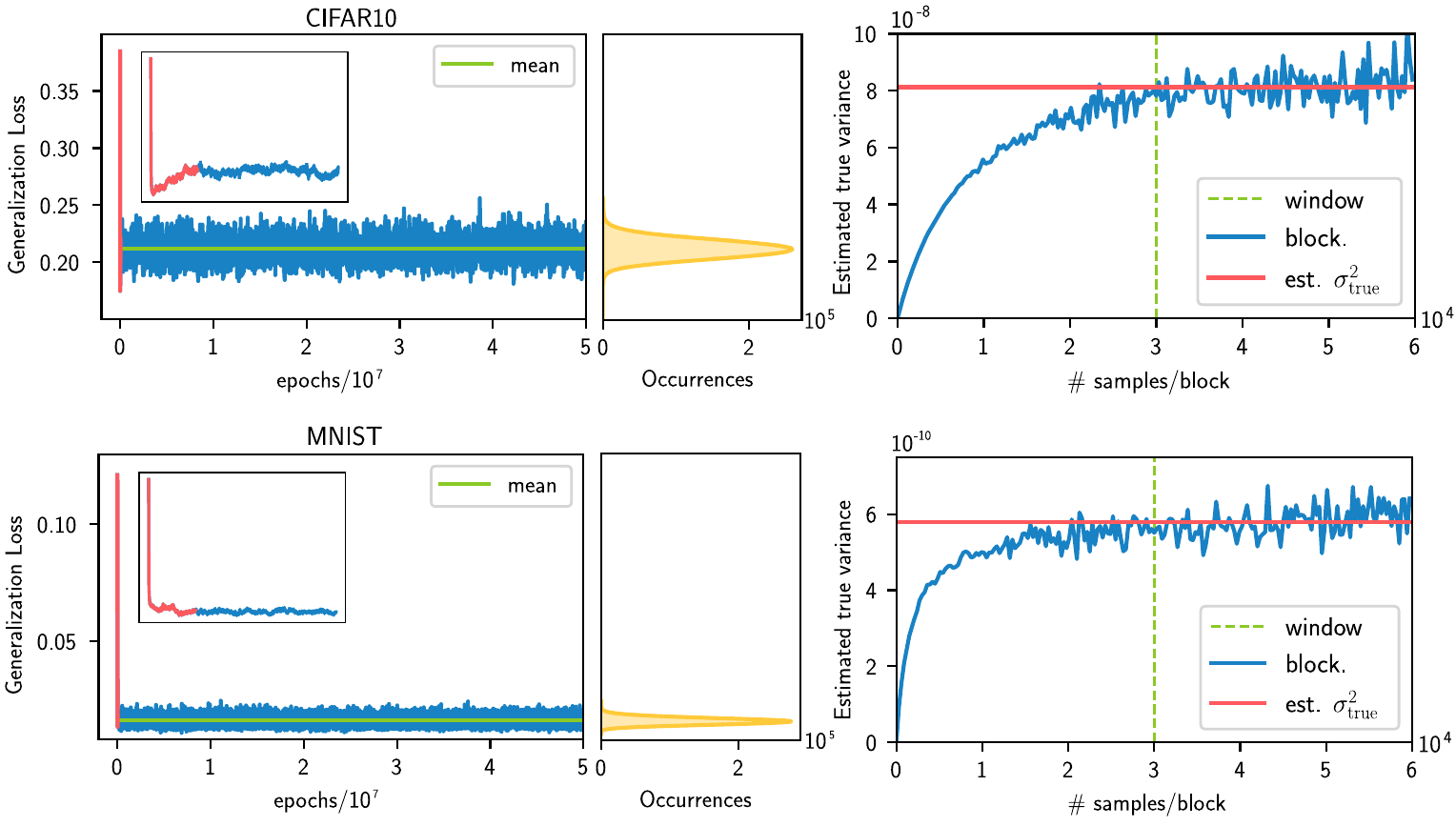}
\caption{Analysis of Monte Carlo data for CIFAR10 dataset ($N_0=32$, first row) and MNIST dataset ($N_0=28$, second row) with $N_1 = P = P_{\mathrm{test}} = 1000$, $T=0.01$, $\lambda_0 = \lambda_1 = 1.0$ and $\epsilon = 0.001, 0.0025$ respectively. On the left are shown examples of generated data (in blue the generalization error calculated on the neural network realizations at equilibrium, in red - see the inset - measurements in the thermalization zone). We display in green the central value (estimated only by using the thermalized region). Next to the stochastic signal, the distribution of measurements with their occurrences is shown in yellow (using samples of $5 \cdot 10^6$ measurements). On the right figures are shown the statistical error estimation with blocking method: blue lines are the standard error of the blocked data for different block sizes and red lines are the estimation of the true error (the plateau of the blocking signal). The plateau was estimated as the average of the standard errors for block values greater than the window value (i.e., considering all the data to the right of the green line).}
\label{fig:signal}
\end{figure}

In this section, we report on the estimation of the generalization error in Eq. \eqref{eq:test_loss} for one FC hidden layer neural network. We aim to measure the expectation value
\begin{equation}
    \epsilon_g = \frac{1}{P_{\mathrm{test}}}\sum_{k=1}^{P_{\mathrm{test}}} \braket{\epsilon_\text{g}(x^k,y^k)}_{\beta} \quad \mathrm{where} \quad
    \braket{\epsilon_\text{g}(x^k,y^k)}_{\beta} = \frac{1}{Z} \int \!\mathcal D \theta \left[ y^k - f_{\mathrm{1HL}}(x^k) \right]^2 \,e^{-\beta \mathcal L (\theta)}\,.
\label{supplemental:gen_error_Ptest_1}
\end{equation}
Widely used for numerical simulations in statistical physics \cite{Binder2002} and lattice gauge theories \cite{Rothe2012, Gattringer2010}, basically a Monte Carlo computations simply amounts of substituting the statistical average in the RHS of Eq. \eqref{supplemental:gen_error_Ptest_1} with an average over a temporal evolution (in the simulation time):
\begin{equation}
    \braket{\epsilon_\text{g}(x^k,y^k)}_{\beta} \approx \frac{1}{N_{s}}\sum_{i=1}^{N_{s}} \epsilon_\text{g}(\theta_i; x^k,y^k)
\label{supplemental:time_avg}
\end{equation}
where $N_{s}$ is the number of generated configurations $\theta_i$ and $\epsilon_\text{g}(\theta_i; x^k,y^k)$ is the value of $\left[ y^k - f_{\mathrm{1HL}}(x^k) \right]^2$ on the $i$-th realization. Roughly speaking, the only requirement for temporal histories to approximate the true value in Eq. \eqref{supplemental:gen_error_Ptest_1} is that the stochastic process admits the canonical distribution $P[\theta] = e^{-\beta\mathcal{L}(\theta)}/Z$ as stationary solution. Under very general assumptions, discrepancies between the true value and the average over the importance sampling are expected to scale as $1/\sqrt{N_{s}}$.

\subsection{Langevin Monte Carlo}

In this work the Langevin Monte Carlo algorithm was used. Indeed configurations that evolve according to the Langevin equation are associated with probability distributions that satisfy the Fokker-Planck equation, allowing for correct sampling from the canonical distribution in the limit of stationary process. The stochastic equation reads:
\begin{equation}
    \dot{\theta}^j = - \frac{\partial \mathcal{L}}{\partial \theta^j} + \sqrt{2T} \eta^j (t)\,.
\label{supplemental:langevin}
\end{equation}
In Eq. \eqref{supplemental:langevin} the last term is a Gaussian withe noise term, normalized so that the following holds
\begin{equation}
    \braket{\eta^i(t_1)}_{\eta} = 0 \quad\quad \braket{\eta^i(t_1)\eta^j(t_2)}_{\eta} = \delta_{ij}\delta(t_1 - t_2)
\end{equation}
where $\braket{\dots}_{\eta}$ means an average over all possible noise realizations.
In the practical implementation, Eq. \eqref{supplemental:langevin} can be used to generate configurations if a finite integration step $\epsilon$ is introduced and the equation is integrated numerically, namely:
\begin{equation}
    \theta^j(t+\epsilon) = \theta^j(t) - \epsilon \frac{\partial \mathcal{L}}{\partial \theta^j(t)} + \sqrt{2T\epsilon} \eta^j (t)
\label{supplemental:langevin_discretized}
\end{equation}
Having introduced a finite integration step, Eq. \eqref{supplemental:langevin_discretized} no longer samples correctly from the Boltzmann distribution but will contain systematic errors (see discussions in section \ref{supplemental:finite_epsilon_effects} for more details). The corrections to the distribution are $\mathcal{O}(\epsilon^n)$, where $n$ is the degree of the integrator used to integrate the Langevin equation ($n=1$ in our case, having opted for the simplest choice, namely the Euler integrator). Typically, this results in a much more optimal tool compared to the more common Metropolis algorithm, as it proposes non-local updates \cite{Barbu2020} (with the additional cost, however, to compute all the gradients of the loss function).

\subsection{Statistical analysis}

Through Eq. \eqref{supplemental:langevin_discretized}, we are able to generate a chain of configurations of neural network weights. In Fig. \ref{fig:signal} we display the signal for a single Monte Carlo run, both for CIFAR10 and MNIST datasets. We generated $50 \cdot 10^6$ configurations, saving the measurements of the generalization error only once every 10 updates.

The system was initialized by drawing the weight values from a normal distribution with zero mean and unit standard deviation. Typically, these configurations are expected to be far from the typical configurations contributing to the integral in Eq. \eqref{supplemental:gen_error_Ptest_1}, so that initial Monte Carlo updates only serve to bring the system to thermalization. The thermalization regions were identified (a default value of 5000 configurations was set, checked by visual inspection, and increased when necessary) and the data up to this point are excluded from the calculation of the mean and statistical errors. In Fig \ref{fig:signal}, insets are shown which represent examples (in red the thermalization regions and in blue the equilibrium regions) for both datasets used.

For each simulation the estimation of the central value is obtained from the average over all generated configurations at the equilibrium. For the estimation of the true statistical error, the blocking method was used (see Fig. \ref{fig:signal}, where examples for both datasets are reported). The blocking method is a statistical technique used to estimate error in correlated samples. It works by dividing the data into blocks and then averaging within each block, effectively reducing correlation between blocks. As the block size increases, the error estimate (with the naive standard error of the mean) over the blocks becomes more reliable, reaching a plateau when blocks are sufficiently uncorrelated (see \cite{Gattringer2010} for further details). The value of the plateau returns the true statistical error and the integrated autocorrelation time $\tau_{int}$ can be computed (as mentioned before, even though we save only once every $10$ steps, we have always found correlations $\tau_{int} \gg 1$). Some statistical error estimates on the samples were also repeated using the Gamma Function method \cite{Wolff2004}, providing completely consistent results.

\subsection{Finite-$\epsilon$ effects}
\label{supplemental:finite_epsilon_effects}

\begin{figure}[t]
\includegraphics[width=1.\textwidth]{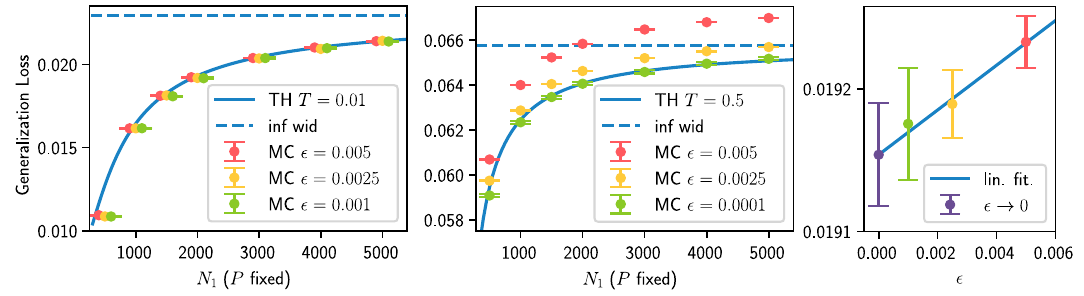}
\caption{Analysis of the finite-$\epsilon$ effects in the case of MNIST dataset ($N_0=28$). The parameters were set to $P = P_{\mathrm{test}} = 1000$, $\lambda_0 = \lambda_1 = 1.0$ and the various temperatures and discrete step values are reported in the legend. The continuous blue line is the prediction of the theory in the proportional limit while the dashed blue line is the prediction of the infinite width theorem. The colored circular markers refer to the neural network simulated at different values of $\epsilon$ (each distinguished by different colors). As can be observed, at low temperatures (left plot), the systematic effects are negligible. With increasing temperature, the error from numerical integration grows (central plot), and it becomes necessary to systematically reduce the value of $\epsilon$, until it aligns with the theory curve for step values of $10^{-4}$ (particularly, too large finite learning rate values appear to even violate the limit predicted by the infinite width theorem). No such discrepancies can be observed considering even more small $\epsilon$. In the first plot, given the very small finite-$\epsilon$ effects, we have implemented a small shift along the $N_1$-axis on the two outer values for easy viewing. Similarly to Fig \ref{fig:lambda}, these values were calculated at the same value of $N_1$, represented by the central point. In the right plot, an example of extrapolation for $\epsilon\to 0$ is shown (the color code is the same as before). All the three finite-$\epsilon$ measurements are practically consistent with the extrapolated value at the statistical precision considered.}
\label{fig:finite_epsilon}
\end{figure}

Removing finite-$\epsilon$ effects that come from the numerical integration of Eq. \eqref{supplemental:langevin_discretized} requires great care. Usually it is very complicated to estimate the size of these systematic errors without some additional steps. In particular, in this work, we observed large systematics at high $T\sim 0.5$, and negligible finite-$\epsilon$ effects at low $T\sim0.01$ (in Fig. \ref{fig:finite_epsilon} we show in details the deviations for the MNIST dataset). 

One option is to insert an additional Metropolis-like acceptance step and use the Langevin algorithm as an optimized update proposal \cite{Barbu2020}. This option is particularly cumbersome as it requires keeping at least two neural network configurations in memory (the new proposal and the old realization). Furthermore it is necessary to fine-tune the $\epsilon$ parameter to achieve an adequate acceptance rate.

A second possibility is to extrapolate the data for $\epsilon\to 0$ \cite{Rothe2012}. If very small step values are considered, these corrections appear as first-order corrections, and linear fits in $\epsilon$ need to be considered. Even in this case, each extrapolation requires extra calculations (data for at least three values of $\epsilon$ are needed).

The last option, which is the one used in this work, is to simulate the system for values of $\epsilon$ so small that no discrepancy can be detected (for example, by comparing the results with an extrapolation or by simulating the system for even smaller $\epsilon$ values). This implies handling very small step values, and therefore large autocorrelation. However, given the large number of configurations at our disposal, the autocorrelations have been observed to still be completely under control. In Fig. \ref{fig:finite_epsilon} (right plot), an example has been reported for the MNIST dataset: the value of the extrapolated generalization error for $\epsilon\to 0$ is completely consistent with the values calculated at finite $\epsilon$ (the errorbars represent one standard deviation). Each value for the generalization error was then calculated for increasingly smaller values of $\epsilon$ until no systematic effect can be observed at fixed statistical precision.

\end{document}